\documentclass[prapplied, reprint,showpacs,  superscriptaddress]{revtex4-1}

\usepackage{color,xcolor}

\usepackage{footnote}
\usepackage[colorinlistoftodos,bordercolor=red!50,backgroundcolor=red!10,linecolor=red!50,textsize=tiny]{todonotes}
\usepackage{changes}

\usepackage[utf8]{inputenc}
\usepackage[T1]{fontenc}
\usepackage[english]{babel}  
\usepackage{times}
\usepackage{bm}

\usepackage[colorlinks=true,linkcolor=blue,citecolor=blue,urlcolor=blue,pdfauthor={ },pdftitle={ },pdfsubject={ },pdfkeywords={ }]{hyperref}
\usepackage{amssymb,amsmath,amsfonts,amsthm}
\usepackage{mathtools}
\usepackage{verbatim}
\usepackage{enumerate}
\usepackage{graphicx}
\usepackage{hyperref}
\usepackage{bbm}
\usepackage{booktabs}
\usepackage{colortbl}

\definecolor{mygray}{gray}{.9}

\usepackage[caption=false]{subfig}

\usepackage{color}
\definecolor{dred}{rgb}{.8,0.2,.2}
\definecolor{ddred}{rgb}{.8,0.5,.5}
\definecolor{dblue}{rgb}{.2,0.2,.8}
\definecolor{dgreen}{rgb}{.2,0.5,.2}

\newcommand{\bra}[1]{\mbox{$\langle #1|$}}
\newcommand{\ket}[1]{\ensuremath{|#1\rangle}}

\newcommand{\be}{\begin{equation}}
\newcommand{\ee}{\end{equation}}
\newcommand{\bea}{\begin{eqnarray}}
\newcommand{\eea}{\end{eqnarray}}

\begin{document}

\title{Improved Quantum State Tomography for the Systems with XX+YY Couplings and Z Readouts}

\author{Tao Xin}
\email{xint@sustech.edu.cn}
\affiliation{Shenzhen Institute for Quantum Science and Engineering, Southern University of Science and Technology, Shenzhen 518055, China}
 \affiliation{Guangdong Provincial Key Laboratory of Quantum Science and Engineering, Southern University of Science and Technology, Shenzhen, 518055, China}
  \affiliation{Shenzhen Key Laboratory of Quantum Science and Engineering, Southern University of Science and Technology, Shenzhen,518055, China}

\begin{abstract}
Quantum device characterization via state tomography plays an important role in both validating quantum hardware and processing quantum information, but it needs the exponential number of the measurements.  For the systems with XX+YY-type couplings and Z readouts, such as superconducting quantum computing (SQC) systems, traditional quantum state tomography (QST) using single-qubit readout operations at least requires $3^n$ measurement settings in reconstructing an $n$-qubit state. In this work, I proposed an improved QST by adding 2-qubit evolutions as the readout operations and obtained an optimal tomographic scheme using the integer programming optimization. I respectively apply the new scheme on SQC systems with the Nearest-Neighbor,  2-Dimensional, and All-to-All connectivities on qubits. It shows that this method can reduce the number of measurements by over 60\% compared with the traditional QST. Besides, comparison with the traditional scheme in the experimental feasibility and robustness against errors were made by numerical simulation. It is found that, the new scheme has good implementability and it can achieve comparable or even better accuracy than the traditional scheme. It is expected that the experimentalist from the related fields can directly utilize the ready-made results for reconstructing quantum states involved in their research.
  \end{abstract}

\maketitle
\section{Introduction}
For an $n$-qubit state, $4^n-1$ unknown coefficients are required to fully characterize the state (considering the normalization). Quantum state tomography (QST) is the processing of determining these coefficients. In the physical implementations, QST is typically realized by measuring the partial information of the state of the system, until the complete density matrix of the system is reconstructed \cite{busch1991informationally}.  On the one hand,  QST is a readout technique that can provide us with the results of quantum tasks performed on quantum devices. On the other hand, QST being a characterization tool is used to benchmark, validate, and develop quantum hardware. It also provides an objective indicator for comparing the performance between different quantum devices. Hence, QST has become an indispensable part of quantum information processing \cite{baur2012benchmarking, lvovsky2009continuous,erhard2019characterizing}. 

Unfortunately,  the number of measurements required by standard QST generally scales exponentially with the system size \cite{cramer2010efficient, lanyon2017efficient}. So far, different quantum platforms are developing more and more controlled qubits towards large-scale quantum computing \cite{gyongyosi2019survey}. On SQC systems,  the teams from Google and IBM have claimed that the number of controlled qubits was up to 53 qubits in the professional laboratory \cite{arute2019quantum} and 16 qubits in open quantum computing cloud \cite{wootton2018repetition, wang201816-qubit}. In the nuclear magnetic resonance (NMR), they already can control up to 12 qubits where 12-qubit quantum pseudo-randomness was realized \cite{li2019experimental}. The photonic quantum computing and ion trap have achieved the control of up to 10 and 14 qubits \cite{wang2017high-efficiency,monz201114-qubit}, respectively.  So consuming millions of measurements and large amounts of resources are inevitable if we want to perform standard QST on these systems in the future. For instance, for reconstructing an 8-qubit physical quantum state on the ion trap, the experimental measurements and classical post-processing cost around one week \cite{haffner2005scalable, klimov2008optimal}. In Ref. \cite{song201710-qubit}, they spent over two days to estimate a 10-qubit GHZ state on the SQC system, where over $5.9\times 10^{4}$ measurement settings were used and $8\times 10^{4}$ experiments for each measurement setting were repeated. Obviously, the exponentially increasing complexity in QST already hinders the applications of large-scale quantum processors in exploring quantum advantages. 

The research for reducing the complexity of QST is divided into two categories, one is for general states, and the other is for constrained states. For a general quantum state in Hilbert space, it is almost impossible to develop a tomographic scheme with the polynomial complexity, but it may be designed and optimized for achieving the fewer measurements. In current quantum platforms, a tomographic process usually contains a series of different measurement settings and performs repeated experiments in each measurement setting. Partial information of the state is extracted from each measurement. When the collected information from such experiments covers all the unknown coefficients of the state,  the density matrices of system can be uniquely reconstructed. However, the obtained information from different measurement settings often overlaps each other. In such case, the key to reducing the complexity of QST is to remove redundant measurements as more as possible by optimizing the tomographic scheme. The related progress have been made in NMR and optical systems \cite{li2017optimal, leskowitz2004state}.  For some constrained quantum states, such as the evolution dynamic and ground states of $k$-local Hamiltonian, there indeed exist some tomographic methods with the polynomial complexity due to the prior information about the state. In the past decades, there have been developments for improving the efficiency of QST using some techniques, such as QST via 2-body reduced density matrices \cite{linden2002parts, xin2017quantum}, compressed sensing \cite{riofrio2017experimental, ahn2019adaptive}, machine learning \cite{torlai2019integrating, palmieri2019experimental, xin2019local-measurement-based}, matrix product states \cite{cramer2010efficient, lanyon2017efficient}, and parameterized quantum circuits \cite{PhysRevApplied, PhysRevA}.

Like the above, in a tomographic scheme for general states, there are often overlaps between the partial information obtained from different measurement settings, such that the measured information is usually overcomplete and some measurement settings are redundant. A natural question is how to find a tomographic scheme with the minimum number of measurement settings that still uniquely reconstruct the density matrices of systems. In this work, I investigate this problem by considering 2-qubit evolutions as the options of the readout operations for the systems with XX+YY couplings and Z readouts. I demonstrate how to find an optimal tomography using integer programming optimization. For the traditional scheme where only single-qubit rotations are chosen as the readout operations \cite{song201710-qubit}, I rigorously prove that at least $3^n$ measurement settings are necessary for performing QST and it is impossible to further reduce the number of measurement settings. Besides, I apply the new scheme on SQC systems and find that an optimal tomography can be given and the number of measurement settings can be reduced by over 60\% when considering 2-qubit evolutions as the readout operations, where 2-qubit evolutions refer to the free evolutions of the natural interactions between two qubits and they are easily implemented using the standard techniques on SQC systems. 

The paper is organized as follows. In Section \ref{sec2}, I mainly recall the traditional QST for the systems with XX+YY couplings and Z readouts and describe the new QST scheme using 2-qubit evolutions. Then I demonstrate how to find the optimal QST using the language of integer programming optimization. Last, a 2-qubit example is presented. In Section \ref{sec3}, I give the detailed results of the optimal tomographic schemes for SQC systems with three different configurations, including the Nearest-Neighbor (NN),  2-Dimensional (2D), and All-to-All (AA) connectivities. In Section \ref{sec4}, I discuss in detail the features of the new scheme in terms of feasibility, accuracy, and scalability. Finally, the summary and outlook are presented in Section \ref{sec5}.

\section{Tomographic scheme}\label{sec2}
\subsection{Problem Description}
A general $n$-qubit state is fully described by a $2^n\times 2^n$ density matrix $\rho$. Due to the completeness and tracelessness of the Pauli matrices, $\rho$ is usually decomposed into the linear combination of the complete Pauli basis with different weights,
\begin{equation}\label{rho}
\rho=\sum^{4^n}_{i=1}\mu_i\mathcal{P}_i, ~\mu_i=\frac{1}{2^n}\text{Tr}(\rho\mathcal{P}_i), ~\mu_1=\frac{1}{2^n}.
\end{equation}
Here, $\mathcal{P}_i$ is a product operator of different Pauli matrices belonging to different qubits. Namely, $\mathcal{P}_i\in \mathcal{S}_{P}: \{I, \sigma_x, \sigma_y, \sigma_z\}^{\otimes n}$, with the Pauli matrices $\sigma_x, \sigma_y, \sigma_z$. $\mu_i$ is the expectation value of the operator $\mathcal{P}_i$, which is the unknown coefficient to be determined. Equation (\ref{rho}) is a convenient form to measure the observables and design the tomographic scheme. In different quantum platforms,  $\mu_i$'s are often obtained by the different principles of measurement. For instance, as the ensemble, NMR can directly measure the expectation value $\mu_i$ from experimental spectra \cite{vandersypen2005nmr}. In other platforms, such as SQC and optical systems, the probability distribution in the eigenstates of the operator $\mathcal{P}_i$ is firstly created by repeating a large number of experiments and then $\mu_i$ is computed \cite{krantz2016single, krantz2019quantum}. 

\begin{figure}[htp]
\centering  
\includegraphics[width=1\linewidth]{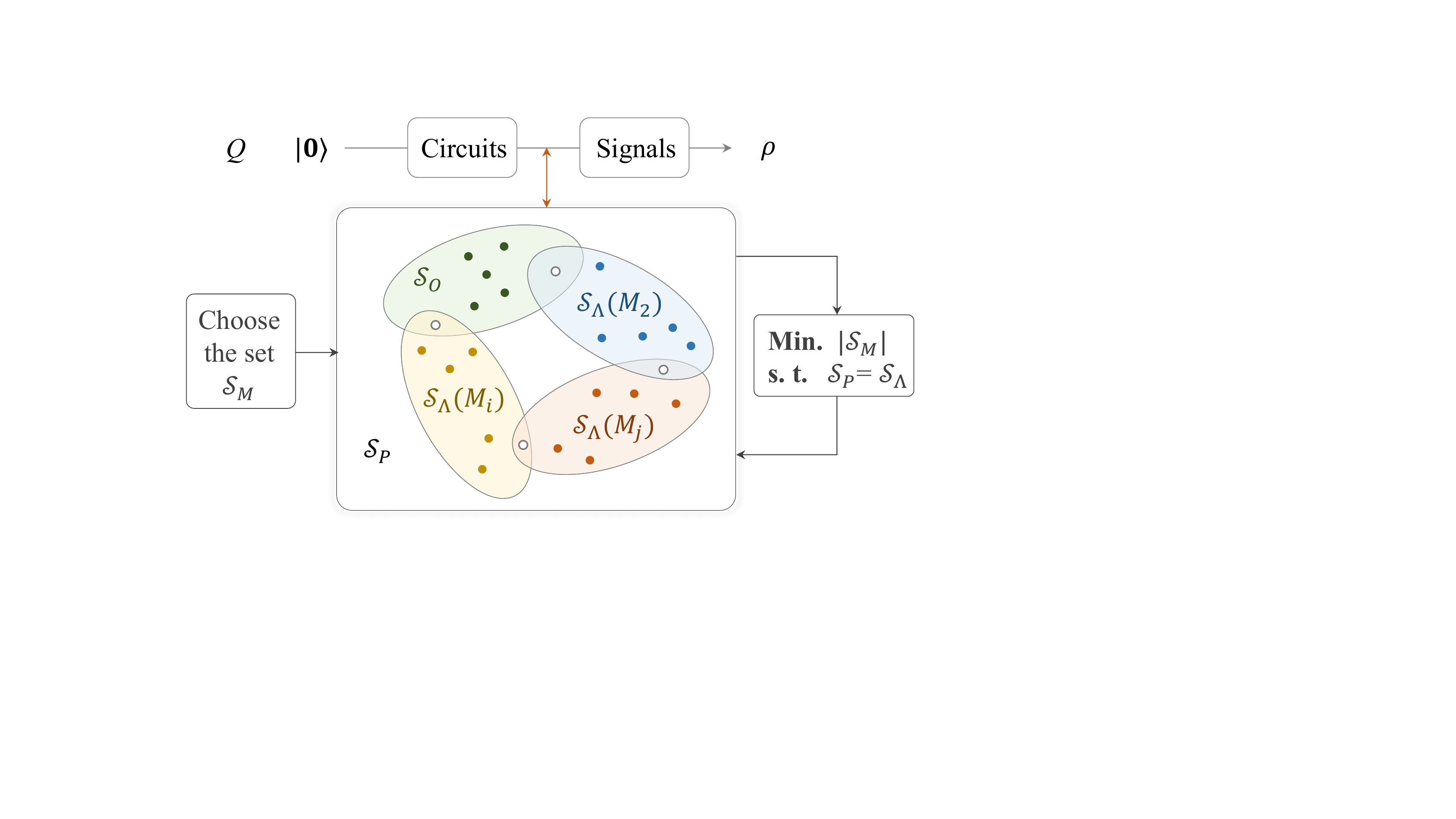}  
\caption{The workflow and schematic diagram for designing a tomographic scheme and reconstructing a state in quantum platforms. The set of the experimentally measurable operators $\mathcal{S}_{O}$ is transferred to the set $\mathcal{S}_{\Lambda}(M_j)$ under the measurement setting $M_j\in \mathcal{S}_{M}$. Usually, some overlap operators between them (labeled by the white circles) exist. Designing an optimal tomographic scheme is to minimize the number of the set $\mathcal{S}_{M}$ (denoted by $|\mathcal{S}_{M}|$) when the set $\mathcal{S}_{\Lambda}\equiv\cup_j\mathcal{S}_{\Lambda}(M_j)$ covers $\mathcal{S}_P$. Then, the optimized measurement set $\mathcal{S}_{M}$ can be implemented after performing circuits and before acquiring signals, such that a state $\rho$ is reconstructed with the minimum number of measurement settings. }
\label{protocol}
\end{figure}

There are $4^n-1$ unknown coefficients for a state $\rho$ to be reconstructed. It is impossible to determine all the coefficients by one measurement setting. Generally speaking, a measurement set including different measurement settings is required in a tomographic experiment. Here, a measurement setting refers to a specific configuration of the detectors (e.g.  optical system \cite{PhysRevxx}) or an applied unitary readout operation after finishing quantum circuits and before acquiring signals (e.g. SQC \cite{krantz2019quantum} and NMR systems \cite{cory1997ensemble}). Let the set of the experimentally measurable operators is $\mathcal{S}_{O}=\{O_1, O_2, ..., O_i, ...\}$ without any readout operations, and the measurement set including available readout operations is denoted by $\mathcal{S}_{M}=\{M_1, M_2, ..., M_j, ...\}$. $O_i\in \mathcal{S}_{P}$ is a product operator of Pauli matrices and $M_j$ is a unitary operation of Clifford group. Under a measurement setting $M_j\in \mathcal{S}_{M}$, the set of experimentally measurable operators will be equivalent to $\mathcal{S}_{\Lambda}(M_j)=\{\Lambda_1, \Lambda_2, ..., \Lambda_i, ...\}$, with the element $\Lambda_i=M_j^\dagger O_iM_j$ due to $\text{Tr}(M_j\rho M_j^\dagger O_i)=\text{Tr}(\rho M_j^\dagger O_iM_j)$. Obviously, $\mathcal{S}_{\Lambda}(M_j)\subseteq \mathcal{S}_{P}$ for all the measurement settings $M_j \in \mathcal{S}_{M}$. When  the set $\mathcal{S}_{P}$ is completely covered by the set $\mathcal{S}_{\Lambda}\equiv\cup_j\mathcal{S}_{\Lambda}(M_j)$, the state $\rho$ will be fully reconstructed. Therefore, there should exist a set $\mathcal{S}_{M}$ with the minimum number of elements under the condition that $\mathcal{S}_{P}$ is covered by $\mathcal{S}_{\Lambda}$. Figure \ref{protocol} presents the schematic diagram of the above logic.

Hence, designing and optimizing a tomographic scheme include the following two steps: (i) How to choose the measurement set $\mathcal{S}_{M}$ consisting of experimentally available readout operations. It is the premise for us to reduce the number of measurement settings required for QST. (ii) Under the chosen measurement set $\mathcal{S}_{M}$, how to find a subset $\{M\}_s\subseteq \mathcal{S}_{M}$ with the minimum number of elements that can still realize QST. In the following,  I will use 2-qubit evolutions as readout operations and explore the above questions for the systems with XX+YY couplings and Z readouts.

\subsection{Traditional QST Scheme }\label{sec3b}

For the systems with Z readouts, the set of the measurable operators $\mathcal{S}_{O}$ includes $2^n$ elements and each operator $O_i $ belongs to the set $\{I, \sigma_z\}^{\otimes n}$. It means that $2^n$ unknown coefficients in Eq. (\ref{rho}) can be determined by one measurement setting. Hence, at least $4^n/2^n=2^n$ measurement settings are required to fully determine a quantum state for arbitrary set $\mathcal{S}_{M}$. It is worth mentioning that the number of measurement settings is multiplied by $2^n$ to yield the actual number of necessary measurements. Considering that one measurement setting creates the same number of unknowns, the factor $2^n$ is not counted in the following and the values refer to the number of measurement settings instead of the total number of measurements.  Complete QST is also possible on the systems with the access-limited measurements \cite{PhysRevLett.124.010405, arenz2020drawing}, for instance, QST for entangled states was realized by only measuring the first qubit in Ref. \cite{PhysRevLett.124.010405}.

As mentioned before, how to choose the set $\mathcal{S}_{M}$ is the key to reducing the number of measurement settings. In the traditional scheme, the applied unitary readout operations are usually selected from the set $\{\mathcal{I}, \mathcal{R}_x, \mathcal{R}_y\}$ acting on the single qubit, where $\mathcal{I}$ is an identity operator, and $\mathcal{R}_x\equiv\text{exp}(-i\pi/4\sigma_x)$ and $\mathcal{R}_y\equiv\text{exp}(-i\pi/4\sigma_y)$ are the $\pi/2$ rotations around the $x$ and $y$ axes, respectively. This corresponds to the measurement set $\mathcal{S}_{M}=\{\mathcal{I}, \mathcal{R}_x, \mathcal{R}_y\}^{\otimes n}$ that includes $3^n$ of elements. This kind of measurement set $\mathcal{S}_{M}$ has been used on the systems with Z readouts, such as SQC systems \cite{naghiloo2019introduction}. However, I find that $3^n$ measurement settings are necessary to fully reconstruct a quantum state under such a set $\mathcal{S}_{M}$, and there does not exist a smaller subset $\{M\}_s\subseteq \mathcal{S}_{M}$ that can achieve the reconstruction of the states. Next, I present the proof for this employing the recursion theory.

{\it{Lemma 1}}-. $3^n$ is the lower bound of the number of measurement settings required to reconstruct a quantum state under  $\mathcal{S}_{M}=\{\mathcal{I}, \mathcal{R}_x, \mathcal{R}_y\}^{\otimes n}$.

{\it{Proof}}-. The set $\mathcal{S}_{P}$ is firstly divided into $(n+1)$ subsets $\mathcal{S}^{(k)}_{P}$. Here, $k$ is the number of Pauli matrices $\sigma_x$ and $\sigma_y$ in the operator $\mathcal{P}_i\in \mathcal{S}_{P}$.\\
(i) For the operator $\mathcal{P}_i \in \mathcal{S}^{(0)}_{P}=\{I, \sigma_z\}^{\otimes n}$, one measurement setting $\mathcal{I}^{\otimes n}$ is enough to measure $\mathcal{P}_i$'s. \\
(ii) For the operator $\mathcal{P}_i \in \mathcal{S}^{(1)}_{P}$, there is one Pauli matrix $\sigma_x$ or $\sigma_y$ in $\mathcal{P}_i$, and $2C^1_n$ measurement settings where only one qubit occupies $\mathcal{R}_x$ or $\mathcal{R}_y$ and the rest qubits occupy $\mathcal{I}$'s are needed. For instance, the measurement setting $\mathcal{R}^1_x\mathcal{I}^2$ is used to measure the operators $\sigma_yI$ and $\sigma_y\sigma_z$ for a 2-qubit system. \\
(iii) For the operator $\mathcal{P}_i \in \mathcal{S}^{(2)}_{P}$, there are two Pauli matrices $\sigma_x$ or $\sigma_y$ in $\mathcal{P}_i$, and we need $2^2C^2_n$  measurement settings where only two qubits occupy $\mathcal{R}_x$ or $\mathcal{R}_y$ and the rest qubits occupy $\mathcal{I}$'s. For example, the measurement setting $\mathcal{R}^1_x\mathcal{I}^2\mathcal{R}^3_y$ is used to measure the operators $\sigma_y I \sigma_x$ and $\sigma_y\sigma_z\sigma_x$ for a 3-qubit system. \\
(iv) By that analogy, $2^nC^n_n$ measurement settings are needed to measure the operator $\mathcal{P}_i \in \mathcal{S}^{(n)}_{P}$. Then, the minimum number of required measurement settings for QST is 
\begin{equation}\label{ss}
|\mathcal{S}_{M}|_{\text{min}}=1+\sum^n_{k=1}2^kC^k_n=3^n.
\end{equation}\label{ss}

 In the following, the above process for reconstructing a quantum state is referred as the traditional QST for the comparison with the new scheme I proposed in this work. 

\subsection{The New QST Scheme}\label{sec2c}

\textit{New measurement settings.}- Section \ref{sec3b} shows that there is no a tomographic scheme that can achieve QST with the fewer measurement settings than $3^n$, under the set $\mathcal{S}_{M}=\{\mathcal{I}, \mathcal{R}_x, \mathcal{R}_y\}^{\otimes n}$. In principle, any element from the Clifford group can be chosen as the unitary readout operations, because the operator $M^\dagger \mathcal{P}_iM$ also belongs to the set $\mathcal{S}_{P}$ under the Clifford operation $M$ \cite{PhysRevcli}. However, it is impractical to consider the entire Clifford group due to its huge size. To find a more efficient QST method, I additionally consider two types of 2-qubit evolutions as the options of the measurement settings, apart from the single-qubit readout operations $\mathcal{I}, \mathcal{R}_x$, and $\mathcal{R}_y$. They are 
\begin{align}
\mathcal{YY}^{(kl)}&\equiv\text{exp}\left(-i\pi/4\sigma^k_y\sigma^l_y\right), \\
\mathcal{XY}^{(kl)}&\equiv\text{exp}\left(-i\pi/4\sigma^k_x\sigma^l_y\right).
\end{align}
$\mathcal{YY}^{(kl)}$ and $\mathcal{XY}^{(kl)}$ are 2-qubit operations between the $k$-th and $l$-th qubits. For the systems with XX+YY-type couplings whose Hamiltonian is $\mathcal{H}_{\text{int}}=g_{kl}(\sigma^k_x\sigma^l_x+\sigma^k_y\sigma^l_y)$, $\mathcal{YY}^{(kl)}$ and $\mathcal{XY}^{(kl)}$ can be easily implemented with the assist of single-qubit rotation pulses and the free coupling evolutions, 
\begin{align}\nonumber
\mathcal{YY}^{(kl)}&=\text{exp}(-i\mathcal{H}_{\text{int}}\tau)\mathcal{R}^k_y(\pi)\text{exp}(-i\mathcal{H}_{\text{int}}\tau)\mathcal{R}^k_y(-\pi), \\
\mathcal{XY}^{(kl)}&=\mathcal{R}^k_z(-\pi/2)\mathcal{YY}^{(kl)}\mathcal{R}^k_z(\pi/2).
\label{yyxy}
\end{align}
$\tau$ is the free evolution time with the value of $\tau=\pi/8g_{kl}$.  I will
 further discuss the experimental feasibility of these two 2-qubit operations in Section \ref{sec4}. 
 
 Then a new measurement set $\mathcal{S}^\text{new}_{M}=\mathcal{S}_{M}\cup\mathcal{S}^{\text{a}}_{M}$ is constructed, including the previous single-qubit set $\mathcal{S}_{M}$ and the added two-qubit set $\mathcal{S}^{\text{a}}_{M}$. It is
\begin{align}
\mathcal{S}^{\text{a}}_{M}=\{\mathcal{YY}^{(kl)}, \mathcal{XY}^{(kl)}\}\otimes \{\mathcal{I}, \mathcal{R}_x, \mathcal{R}_y\}^{\otimes n-(kl)}.
\end{align}
Compared with the traditional QST under the set $\mathcal{S}_{M}$, it is hopeful to find a more efficient tomographic scheme with fewer measurement settings than $3^n$ under the set $\mathcal{S}^\text{new}_{M}$. As shown in Table \ref{yy}, it is because the introduction of $\mathcal{YY}^{(kl)}$ and $\mathcal{XY}^{(kl)}$ can measure more non-overlap operators of the set $\mathcal{S}_{P}$ than single-qubit readout operations. 

\begin{table}[!htp]
\caption{The measurable operators when $\mathcal{YY}^{(kl)}$ and $\mathcal{XY}^{(kl)}$ are chosen as the measurement settings for $O_i\in\{I, \sigma_z\}^{\otimes 2}$.} 
\begin{tabular}{ccc}
\toprule[1pt]  
~~The operators $O$'s & ~~~$\mathcal{YY}^\dag O \mathcal{YY}$~~~ &  ~~~$\mathcal{XY}^\dag O \mathcal{XY}$~~~ \\
\hline
$II$ & $II$ & $II$\\
$I\sigma_z$ & -$\sigma_y\sigma_x$ & -$\sigma_x\sigma_x$\\
$\sigma_z I$ & -$\sigma_x\sigma_y$ & $\sigma_y\sigma_y$\\
$\sigma_z\sigma_z$ & $\sigma_z\sigma_z$ & $\sigma_z\sigma_z$\\
\toprule[1pt]  
\end{tabular}
\label{yy}
\end{table} 

\textit{Optimal QST scheme.}- Now, the question is how to find a subset $\{M\}_s\subseteq \mathcal{S}^\text{new}_{M}$ with the number of measurement settings as small as possible under the condition $\mathcal{S}_{\Lambda}= \mathcal{S}_{P}$. 
It is a typical set cover problem \cite{halperin2005the, alon2003the}. Although many famous algorithms are proposed, such as the greedy algorithm,  they do not yield the globally optimal solution \cite{hassin2005a, bar1981linear, young2008greedy}. 
In this work, I transfer it to  the integer programming optimization to find the optimal or approximate-optimal solution for this problem \cite{li2017optimal,  schrijver1998theory}. The detailed process is described in the following steps.

\begin{figure}[h]
\centering  
\includegraphics[width=0.9\linewidth]{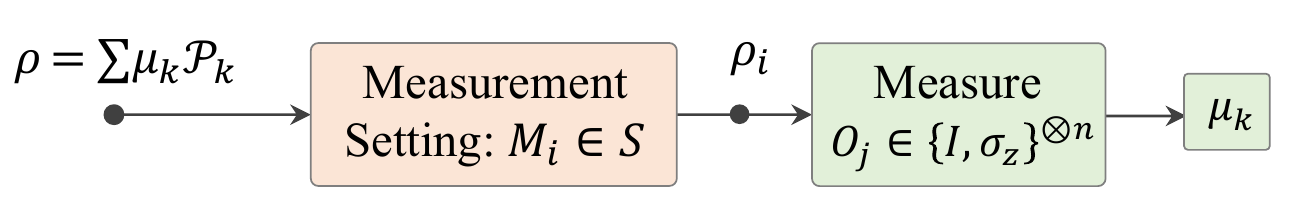}  
\caption{The process of QST for the given measurement settings $M_i\in \mathcal{S}_M^{\text{new}}$ and measurable operators $O_i\in \mathcal{S}_O=\{I, \sigma_z\}^{\otimes n}$.  Different measurement settings are adopted until all unknown parameters $\mu_k$ in $\rho$ are measured.} 
\label{pro}
\end{figure}

$(i)$ \textit{Mathematicize this problem.}- As shown in Fig. \ref{pro}, for an unknown state $\rho=\sum\mu_k\mathcal{P}_k$ with $\mathcal{P}_k \in \mathcal{S}_P=\{I,\sigma_x,\sigma_y,\sigma_z\}^{\otimes n}$, the measurement setting $M_i\in \mathcal{S}_M^{\text{new}}$ is firstly applied on $\rho$ before measurements, and then the expectation value of operators $O_j$ on states $\rho_i=M_i\rho M_i^{\dagger}$ is measured by 
\begin{align}\nonumber
\left \langle O_j \right \rangle&=\text{Tr}(O_j\cdot \rho_i)\\
&=\text{Tr}(O_j\cdot M_i\rho M_i^{\dagger})\\\nonumber
&=\text{Tr}(M_i^{\dagger}O_j M_i \cdot\rho).
\end{align}
So the expectation values of operators $M_i^{\dagger}O_j M_i$ on $\rho$ are obtained by measuring the operator $O_j$ on $\rho_i$. Here, $M_i^{\dagger}O_j M_i$ also belongs to the set $\mathcal{S}_P$. Now, a $|\mathcal{S}^\text{new}_{M}|\times 4^N$ matrix $A$ can be constructed to describe the measurable operators in the measurement settings $M_i$'s, where the column index represents the order of $\mathcal{P}_k$ in the set $\mathcal{S}_P$ and  the row index represents the order of $M_i$ in the set $\mathcal{S}_M^{\text{new}}$. The element $A_{ij}=\text{sgn} \cdot1 (A_{ij}=0)$ means the parameter $\mu_{j}$ can (not) be measured using the measurement setting $M_i$.  Here, sgn$=\pm 1$ is the plus-minus sign between the measured operator  $M_i^{\dagger}O_j M_i$ on $\rho$ and the corresponding operator in the set $\mathcal{S}_P$. So, the first step is to create such a matrix $A$ for the given sets $\mathcal{S}_P, \mathcal{S}_O$, and $\mathcal{S}^\text{new}_{M}$. Table \ref{2q} gives a 2-qubit example.

$(ii)$ \textit{Transfer to the binary integer programming.}- Full QST requires that all the parameters $\mu_k$'s can be measured for a selected subset $\{M\}_s\subseteq \mathcal{S}^\text{new}_{M}$. In other words, each column of matrix $A$ must contain at least one non-zero element for all $M_i$'s in $ \{M\}_s$. Now, there are three problems to be addressed. \\
(1) How to use the formulation to express the selection of the measurement settings in the set $\mathcal{S}^\text{new}_{M}$. \\
(2) How to use the formulation to constraint each column of matrix $A$ to contain at least one non-zero element. \\
(3) How to use the formulation to express the number of selected measurement settings. \\
For the first point, I define a $|\mathcal{S}^\text{new}_{M}|$-dimensional column vector $x$, where each element $x_i$ is zero or one. $x_i=1$ and $x_i=0$ respectively represent the reservation and remove of the $i$-th measurement setting $M_i$ in the set $\mathcal{S}^\text{new}_{M}$. 
For the second point, if the absolute sum of the column of $A$ is more than 1, it means this column contains at least one non-zero element. Mathematically, it is a linear constraint $B^Tx\geqslant 1$ with $B$ being the absolute matrix of $A$ and superscript $T$ being matrix transposition. For the third point, the sum of $x$ is used to express the number of selected measurement settings.  It is $\mathcal{L}(x)=\sum_i x_i$. After defining the one-zero variable $x$, the objective function $\mathcal{L}(x)$, and the constraint condition $B^Tx\geqslant 1$, this problem can be described using the language of the binary integer optimization as follows,
\begin{align}
\text{minimize:} ~~~&\mathcal{L}(x)=\sum_i x_i\\
\text{subject to:} ~~~&B^Tx\geqslant 1, x_i=0~\text{or}~1
\end{align}

\begin{table*}[!htb]
\caption{The measurable parameters $\mu_k$'s under the different measurement settings $M_i$'s. It constructs a sparse matrix $A_{11\times 16}$.}. 
\centering 
\begin{tabular}{|c|c|c|c|c|c|c|c|c|c|c|c|c|c|c|c|c|}

\hline
$M_i$ & ~$\mu_1$~ & ~$\mu_2$~ & ~$\mu_3$~ &~$\mu_4$~ & ~$\mu_5$~ & ~$\mu_6$~ & ~$\mu_7$~ & ~$\mu_8$~ & ~$\mu_9$~  & ~$\mu_{10}$~ & ~$\mu_{11}$~ & ~$\mu_{12}$~ & ~$\mu_{13}$~ & ~$\mu_{14}$~ & ~$\mu_{15}$~ & ~$\mu_{16}$~\\
\hline
$\mathcal{I}\mathcal{I}$ & 1 & 0 & 0 & 1 & 0 & 0 & 0 & 0 & 0  & 0 & 0 & 0 & 1 & 0 & 0 & 1\\
\hline
\rowcolor{mygray}
$\mathcal{R}^2_x$ & 1 & 0 & 1 & 0 & 0 & 0 & 0 & 0 & 0  & 0 & 0 & 0 & 1 & 0 & 1 & 0\\
\hline
\rowcolor{mygray}
$\mathcal{R}^2_y$ & 1 & -1 & 0 & 0 & 0 & 0 & 0 & 0 & 0  & 0 & 0 & 0 & 1 & -1 & 0 & 0\\
\hline
\rowcolor{mygray}
$\mathcal{R}^1_x$ & 1 & 0 & 0 & 1 & 0 & 0 & 0 & 0 & 1  & 0 & 0 & 1 & 0 & 0 & 0 & 0\\
\hline
$\mathcal{R}^1_x\mathcal{R}^2_x$ & 1 & 0 & 1 & 0 & 0 & 0 & 0 & 0 & 1  & 0 & 1 & 0 & 0 & 0 & 0 & 0\\
\hline
$\mathcal{R}^1_x\mathcal{R}^2_y$ & 1 & -1 & 0 & 0 & 0 & 0 & 0 & 0 & 1  & -1 & 0 & 0 & 0 & 0 & 0 & 0\\
\hline
\rowcolor{mygray}
$\mathcal{R}^1_y$ & 1 & 0 & 0 & 1 & -1 & 0 & 0 & -1 & 0  & 0 & 0 & 0 & 0 & 0 & 0 & 0\\
\hline
$\mathcal{R}^1_y\mathcal{R}^2_x$ & 1 & 0 & 1 & 0 & -1 & 0 & -1 & 0 & 0  & 0 & 0 & 0 & 0 & 0 & 0 & 0\\
\hline
$\mathcal{R}^1_y\mathcal{R}^2_y$ & 1 & -1 & 0 & 0 & -1 & 1 & 0 & 0 & 0  & 0 & 0 & 0 & 0 & 0 & 0 & 0\\
\hline
\rowcolor{mygray}
$\mathcal{YY}$ & 1 & 0 & 0 & 0 & 0 & 0 & -1 & 0 & 0  & -1 & 0 & 0 & 0 & 0 & 0 & 1\\
\hline
\rowcolor{mygray}
$\mathcal{XY}$ & 1 & 0 & 0 & 0 & 0 & -1 & 0 & 0 & 0  & 0 & 1 & 0 & 0 & 0 & 0 & 1\\
\hline
\end{tabular}
\label{2q}
\end{table*}

$(iii)$ \textit{Solve this problem.}- The above is the binary linear programming optimization. It belongs to typical integer linear programming. So far,  a large of algorithms are proposed to solve the integer linear programming optimization \cite{chenapplied}. There is also a large collection of open or free-academic solvers that can be used to solve the integer linear programming, such as \href{http://lpsolve.sourceforge.net/5.5/}{lp\_solve}, \href{https://www.scipopt.org/index.php}{SCIP}, \href{https://pypi.org/project/mip/}{MIP}, \href{https://www.gurobi.com}{Gurobi}, and \href{https://www.mosek.com}{Mosek} (Hidden links). These solvers are callable from the common languages, including Python and MATLAB. Besides, there is also a built-in function \href{https://ww2.mathworks.cn/help/optim/ug/intlinprog.html}{intlinprog} in MATLAB language to solve the integer linear programming. It is worth emphasizing that Gurobi and Mosek are commercial solvers but the use-free versions can be obtained for academic purposes.

\subsection{An Example for Reconstructing 2-qubit States}\label{sec2d}
As shown in Fig. \ref{pro}, a 2-qubit state $\rho_0$ to be reconstructed can be decomposed into the linear combination of the Pauli basis. It is $\rho_0=\sum_{k=1}^{16}\mu_k\mathcal{P}_k$ with $\mu_1=0.25$. Here,
\begin{align}\nonumber
\mathcal{P}_k\in &\{II,I\sigma_x,I\sigma_y,I\sigma_z,\sigma_xI,\sigma_x\sigma_x,\sigma_x\sigma_y,\sigma_x\sigma_z,\\& \sigma_yI,\sigma_y\sigma_x,\sigma_y\sigma_y,\sigma_y\sigma_z,\sigma_zI,\sigma_z\sigma_x,\sigma_z\sigma_y,\sigma_z\sigma_z\}.
\end{align}
$\mu_k$'s are unknown parameters to be measured. On the systems with Z readouts, one can only measure the expectation value of operator $O_j\in \mathcal{S}_{O}=\{II,I\sigma_z,\sigma_zI,\sigma_z\sigma_z\}$. It is $\mu_1, \mu_4, \mu_{13}$, and $\mu_{16}$. To obtain other parameters, we need to apply some measurement settings $M_i$ on $\rho_0$ before measurements, to transfer the undetectable operators to the detectable ones.

In the proposed scheme, the available measurement settings
\begin{align}\nonumber
M_i\in S_M^{\text{new}}=\{\mathcal{I}, \mathcal{R}_x, \mathcal{R}_y\}^{\otimes 2}\cup \{\mathcal{YY}^{(12)}, \mathcal{XY}^{(12)}\}.
\end{align}
For instance, when $M_i=\mathcal{R}^2_x$,  the following operator can be measured, 
\begin{align}\nonumber
M_i^{\dagger}O_j M_i\in \mathcal{S}_{\Lambda}(M_j)=\{II, I\sigma_y, \sigma_zI, \sigma_z\sigma_y\}.
\end{align} 
It will yield $\mu_1, \mu_3, \mu_{13}$, and $\mu_{15}$ with,
\begin{align}\nonumber
\mu_1=\frac{\left \langle O_1 \right \rangle}{2^2}, \mu_3=\frac{\left \langle O_2 \right \rangle}{2^2}, \mu_{13}=\frac{\left \langle O_3 \right \rangle}{2^2}, \mu_{15}=\frac{\left \langle O_4 \right \rangle}{2^2}.
\end{align} 
Similarly, when $M_i=\mathcal{YY}^{(12)}$,  one can measure 
\begin{align}\nonumber
M_i^{\dagger}O_j M_i\in \mathcal{S}_{\Lambda}(M_j)= \{II, -\sigma_y\sigma_x, -\sigma_x\sigma_y, \sigma_z\sigma_z\}.
\end{align} 
It will yield $\mu_1, \mu_{10}, \mu_{7}$, and $\mu_{16}$ with,  
\begin{align}\nonumber
\mu_1=\frac{\left \langle O_1 \right \rangle}{2^2}, \mu_{10}=-\frac{\left \langle O_2 \right \rangle}{2^2}, \mu_{7}=-\frac{\left \langle O_3 \right \rangle}{2^2}, \mu_{16}=\frac{\left \langle O_4 \right \rangle}{2^2}.
\end{align} 
Table \ref{2q} presents the measurable parameters under the different measurement settings $M_i$'s. It is an $11\times 16$ sparse matrix (denoted by $A$). So, the question is
how to find a subset $\{M\}_s\subseteq \mathcal{S}^\text{new}_{M}$ with the number of measurement settings as small as possible that can achieve the measurement of all $\mu_j$'s. According to the optimization process described in Section \ref{sec2c}, I use the built-in function $\texttt{intlinprog}$ in MATLAB software to find the optimal solution $x^t=[0, 1, 1, 1, 0, 0, 1, 0, 0, 1, 1]^T$, and the corresponding measurement settings are shown by the gray rows in Table \ref{2q}. Simple verification shows that the new scheme needs 6 measurement settings including $M_2, M_3, M_4, M_7, M_{10}$,and $M_{11}$ to fully reconstruct $\rho_0$, while 9 measurement settings including $M_1$ to $M_9$ are necessary in the traditional scheme.

\section{The applications and results on SQC systems}\label{sec3}
In this section, I apply the new scheme on SQC systems and present the results of optimizing the tomographic scheme, for three common configurations including the AA,  NN, and  2D connectivities between qubits. 


\begin{figure}[htp]
\centering  
\includegraphics[width=1\linewidth]{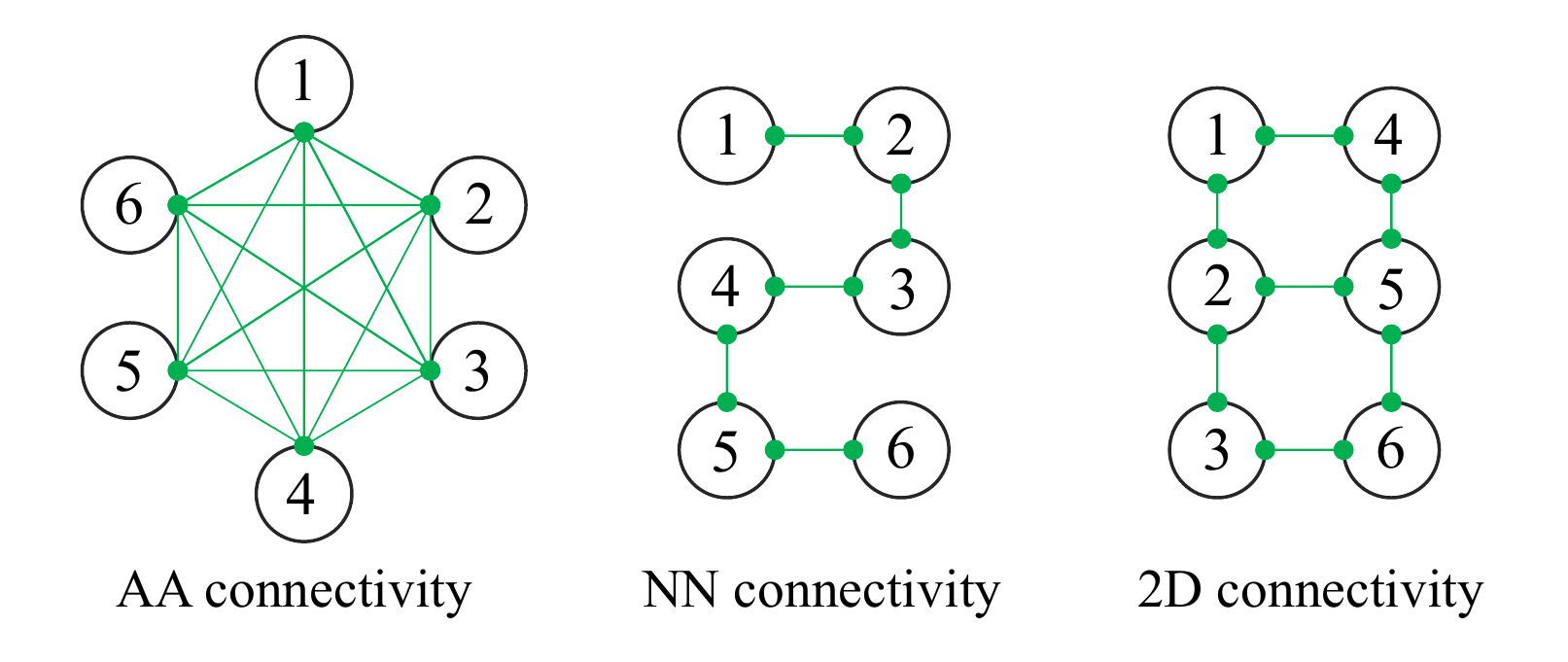}  
\caption{The three common configurations on SQC systems by taking 6-qubit systems as example.}
\label{nn}
\end{figure}

Today, the most used readout method on SQS systems is the so-called dispersive readout \cite{naghiloo2019introduction, Physxxy,PhysRevcc}, where each qubit (as quantum system) couples with a readout resonator (as the detector). The Hamiltonian between the qubits and readout resonator can be described by the Jaynes–Cummings model as follows \cite{shore1993the},
\begin{equation}\label{Hjc}
\mathcal{H}_{\text{JC}}=\frac{\omega_0}{2}\sigma_z+g_{0r}(\sigma_+a+\sigma_-a^\dag)+\omega_r(a^\dag a+\frac{1}{2}).
\end{equation}
Here, $\omega_0$ and $\omega_r$ are the frequencies of the qubit and readout resonator, respectively. $g_{0r}$ is the coupling between the qubit and readout resonator. When the qubit is far detuned from the readout resonator with $\bigtriangleup=|\omega_0-\omega_r|\gg g_{0r}$, in the dispersive approximation \cite{Physvv}, the Hamiltonian can be approximated as
\begin{equation}\label{Hd}
\mathcal{H}_{\text{dis}}=\frac{(\omega_0+g^2_{0r}/\bigtriangleup)}{2}\sigma_z+(\omega_r+\frac{g^2_{0r}}{\bigtriangleup}\sigma_z)(a^\dag a+\frac{1}{2}).
\end{equation}\label{Hd}
It is shown that the frequencies of the qubit and readout resonator influence each other, and the states of the qubit lead to a state-dependent frequency shift of the readout resonator. It is $\omega_r+g^2_{0r}/\bigtriangleup$ for the state $\ket{0}$ or $\omega_r-g^2_{0r}/\bigtriangleup$ for the state $\ket{1}$. This feature allows us to obtain state information about the qubit by directly reading the readout resonator. Hence, the probability distributions in the computational basis from $\ket{00....0}$ to $\ket{11....1}$ can be measured by repeating a large number of experiments on SQC systems. If we denote $\mathbb{P}(\ket{\phi_l})$ as the probability distribution in the $l$-th computational basis $\ket{\phi_l}$,  the expectation value of the operator $O_i $ is then,
\begin{equation}\label{Hd}
\left \langle O_i  \right \rangle=\frac{1}{2^n}\sum^{2^n}_{l=1}\text{Tr}(O_i \cdot\ket{\phi_l}\bra{\phi_l})\cdot\mathbb{P}(\ket{\phi_l}).
\end{equation}\label{Hd}
It corresponds to the measurement of the diagonal elements of the density matrix or the expectation values of the operators in the set $\{I, \sigma_z\}^{\otimes n}$.  On SQC systems, the interaction Hamiltonian between the $k$-th and $l$-th qubits is described as $\mathcal{H}_{\text{int}}=g_{kl}(\sigma^k_x\sigma^l_x+\sigma^k_y\sigma^l_y)$ with the coupling  $g_{kl}$ \cite{song201710-qubit, PhysRevcop, PhysRevLettcop, niskanen2007quantum}. So it can use the proposed scheme in this work for QST.

\subsection{The AA Configuration}
It is known that the connectivity between qubits and the feasibility of two-qubit gates between arbitrary two qubits determine the performance and quality of the performed algorithms on the superconducting chips.  Here, I first consider the AA configuration on SQC systems. As shown in Fig. \ref{nn}, it refers to a fully-connected graph arrangement of qubits, which is a perfect structure the scientists prefer to develop. For instance, Ref. \cite{song201710-qubit} made a programable and fully-connected 10-qubit superconducting processor by coupling all the qubits with a bus resonator, in which each qubit can interact with other qubits with the tunable couplings. On such a system, it is available to realize 2-qubit operations $\mathcal{YY}^{(kl)}$ and $\mathcal{XY}^{(kl)}$ between arbitrary $k$-th and $l$-th qubits, see Section \ref{sec4}. Next, I first prove that there exists a more efficient tomographic scheme with fewer measurement settings than traditional QST.

{\it{Lemma 2}}-.  Using the proposed tomographic scheme, the number of measurement settings required to reconstruct a quantum state is at least reduced to $(3^n+2n+1)/2$ from the traditional $3^n$.

{\it{Proof}}-. Similar with the derivation in Section \ref{sec3b}, the set $\mathcal{S}_{P}$ is divided into $\mathcal{S}_{P}=\mathcal{S}^{(0)}_{P}\cup\mathcal{S}^{(1)}_{P}\cup .... \cup \mathcal{S}^{(N-1)}_{P} \cup \mathcal{S}^{(N)}_{P}$. \\
(i) For the operator $\mathcal{P}_i \in \mathcal{S}^{(0)}_{P}=\{I, \sigma_z\}^{\otimes n}$, one measurement setting $\mathcal{I}^{\otimes n}$ is still used to measure $\mathcal{P}_i$. \\
(ii) For the operator $\mathcal{P}_i \in \mathcal{S}^{(1)}_{P}$, we need $2C^1_n$ single-qubit measurement settings where only one qubit occupies $\mathcal{R}_x$ or $\mathcal{R}_y$ and the rest qubits occupy $\mathcal{I}$'s.  \\
(iii) For the operator $\mathcal{P}_i \in \mathcal{S}^{(2)}_{P}$, there are two Pauli matrices $\sigma_x$ or $\sigma_y$ in $\mathcal{P}_i$. and we need $2\cdot C^2_n$  measurement settings where only two qubits occupy $\mathcal{YY}^{(kl)}$ or $\mathcal{XY}^{(kl)}$ and the rest qubits occupy $\mathcal{I}$'s. For instance, the measurement setting $\mathcal{YY}^{(12)}\mathcal{I}^3$ is capable to measure the operators $\sigma_y\sigma_x I$,  $\sigma_y\sigma_x\sigma_z$,  $\sigma_x\sigma_y I$, and  $\sigma_x\sigma_y\sigma_z$ for a 3-qubit system. \\
(iv) For the operator $\mathcal{P}_i \in \mathcal{S}^{(3)}_{P}$, there are three Pauli matrices $\sigma_x$ or $\sigma_y$ in $\mathcal{P}_i$, and we need $2\cdot 2\cdot C^3_n$ measurement settings where two qubits occupy $\mathcal{YY}^{(kl)}$ or $\mathcal{XY}^{(kl)}$, one qubit occupies $\mathcal{R}_x$ or $\mathcal{R}_y$, and the rest qubits occupy $\mathcal{I}$'s. For instance, the measurement setting $\mathcal{YY}^{(12)}\mathcal{I}^3\mathcal{R}^4_x$ is capable to measure the operators $\sigma_y\sigma_xI\sigma_y$,  $\sigma_y\sigma_x\sigma_z\sigma_y$,  $\sigma_x\sigma_yI\sigma_y$, and  $\sigma_x\sigma_y\sigma_z\sigma_y$ for a 4-qubit system. \\
(v) By that analogy, $2\cdot 2^{n-2} \cdot C^n_n$ measurement settings are necessary to measure the operator $\mathcal{P}_i \in \mathcal{S}^{(n)}_{P}$. The first factor 2 is due to two selections from $\mathcal{YY}^{(kl)}$ or $\mathcal{XY}^{(kl)}$ and the second factor 2 means two selections from $\mathcal{R}_x$ or $\mathcal{R}_y$. In all, the number of required measurement settings for QST is 

\begin{equation}\label{ssn}
1+2C^1_n+\sum^n_{k=2}2^{k-1}C^k_n=\frac{3^n+2n+1}{2}.
\end{equation}

\begin{table}[h] 
\caption{The comparison between traditional QST and the proposed tomographic scheme for three common configurations. The values refer to the number of measurement settings required by QST.}. 
\begin{tabular}{ccccccc}
\toprule[1pt]  
~~Qubit number & ~~~~~2~~~~~ &  ~~~~~3~~~~~ & ~~~~~4~~~~~ & ~~~~~5~~~~~ & ~~~~~6~~~~~ &  ~~~~~7~~~\\
\hline
Traditional QST & 9 & 27 & 81 & 243 & 729 & 2187\\
New scheme-AA & 6 & 15 & 35 & 89 & 265 & 780\\
New scheme-NN & 6 & 16 & 39 & 108& 293 & 837\\
New scheme-2D & $\sim$ & $\sim$ & 38 & $\sim$& 284 & $\sim$\\
\toprule[1pt]  
\end{tabular}
\label{com}
\end{table}

\begin{table*}[!htp]
\caption{The optimal tomographic scheme for the AA configuration. Here, examples of the size with 2 to 4 qubits are presented. }. 
\begin{tabular}{cc}
\toprule[1pt]  
Qubit & ~~~Measurement settings~~~ \\
\hline
2 & $\mathcal{R}^2_x$, $\mathcal{R}^2_y$, $\mathcal{R}^1_x$, $\mathcal{R}^1_y$, $\mathcal{YY}^{(12)}$, $\mathcal{XY}^{(12)}$\\
3 & $\mathcal{R}^2_x\mathcal{R}^3_y$, $\mathcal{R}^2_y\mathcal{R}^3_x$, $\mathcal{R}^1_x\mathcal{R}^3_y$, $\mathcal{R}^1_x\mathcal{R}^2_y$, $\mathcal{R}^1_y\mathcal{R}^3_x$, $\mathcal{R}^1_y\mathcal{R}^2_x$, $\mathcal{XY}^{(12)}$, $\mathcal{XY}^{(12)}\mathcal{R}^3_x$, $\mathcal{XY}^{(12)}\mathcal{R}^3_y$, $\mathcal{XY}^{(13)}$, $\mathcal{XY}^{(13)}\mathcal{R}^2_x$, $\mathcal{XY}^{(13)}\mathcal{R}^2_y$, $\mathcal{XY}^{(23)}$, \\
& $\mathcal{R}^1_x\mathcal{XY}^{(23)}$, $\mathcal{R}^2_y\mathcal{XY}^{(23)}$\\
4 & $\mathcal{R}^3_x$, $\mathcal{R}^3_y$, $\mathcal{R}^1_x\mathcal{R}^2_x\mathcal{R}^4_x$, $\mathcal{YY}^{(12)}$, $\mathcal{XY}^{(12)}$, $\mathcal{YY}^{(12)}\mathcal{R}^4_y$, $\mathcal{XY}^{(12)}\mathcal{R}^3_x\mathcal{R}^4_x$, $\mathcal{YY}^{(12)}\mathcal{R}^3_x\mathcal{R}^4_y$, $\mathcal{YY}^{(12)}\mathcal{R}^3_y\mathcal{R}^4_x$, $\mathcal{XY}^{(12)}\mathcal{R}^3_y\mathcal{R}^4_y$, $\mathcal{YY}^{(13)}$, $\mathcal{XY}^{(13)}$, \\
& $\mathcal{YY}^{(13)}\mathcal{R}^4_x$, $\mathcal{XY}^{(13)}\mathcal{R}^4_y$, $\mathcal{XY}^{(13)}\mathcal{R}^2_x$, $\mathcal{YY}^{(13)}\mathcal{R}^2_y$, $\mathcal{YY}^{(14)}$, $\mathcal{XY}^{(14)}$, $\mathcal{YY}^{(14)}\mathcal{R}^2_x\mathcal{R}^3_x$, $\mathcal{XY}^{(14)}\mathcal{R}^2_x\mathcal{R}^3_y$, $\mathcal{YY}^{(14)}\mathcal{R}^2_y$, $\mathcal{XY}^{(14)}\mathcal{R}^2_y\mathcal{R}^3_x$, \\
& $\mathcal{YY}^{(14)}\mathcal{R}^2_y\mathcal{R}^3_y$, $\mathcal{YY}^{(23)}\mathcal{R}^4_x$, $\mathcal{XY}^{(23)}\mathcal{R}^4_y$, $\mathcal{R}^1_x\mathcal{YY}^{(23)}$, $\mathcal{R}^1_y\mathcal{XY}^{(23)}$, $\mathcal{YY}^{(24)}$, $\mathcal{XY}^{(24)}$, $\mathcal{R}^1_x\mathcal{YY}^{(24)}$, $\mathcal{R}^1_y\mathcal{XY}^{(24)}$, \\
& $\mathcal{R}^2_x\mathcal{XY}^{(34)}$, $\mathcal{R}^2_y\mathcal{YY}^{(34)}$,  $\mathcal{R}^1_x\mathcal{XY}^{(34)}$, $\mathcal{R}^1_y\mathcal{YY}^{(34)}$\\
\toprule[1pt]  
\end{tabular}
\label{aaresult}
\end{table*} 

\begin{figure*}[htp]
\centering  
\includegraphics[width=0.8\linewidth]{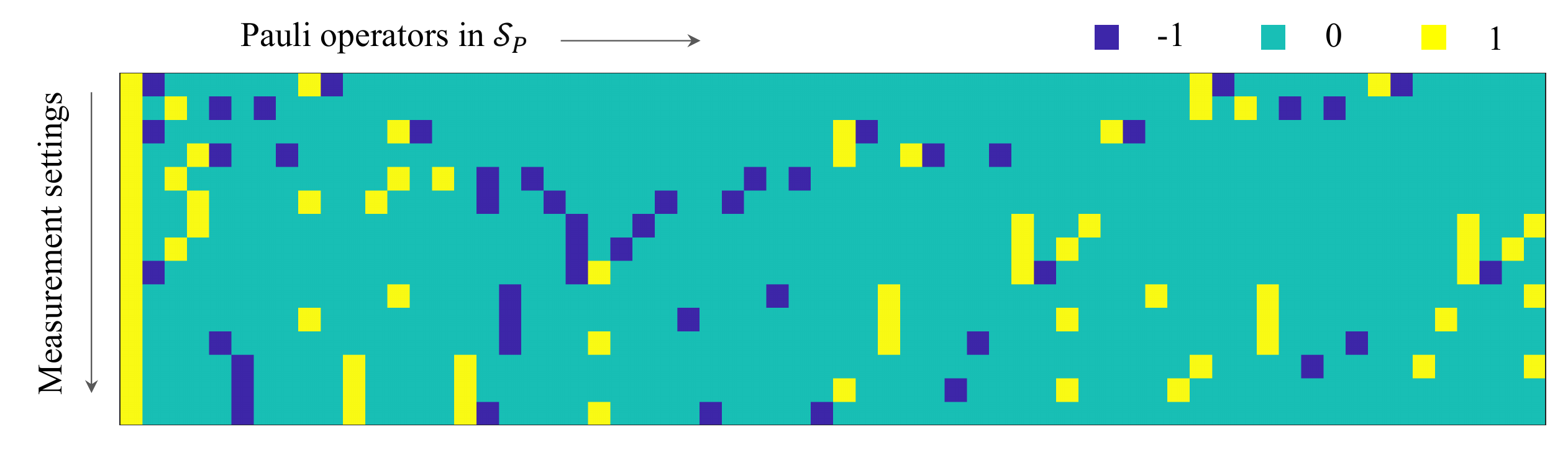}  
\caption{The values of the optimal matrix $A^t$ after removing the redundant measurement settings for a 3-qubit AA configuration. It is a $15\times 64$ matrix, in which the horizontal axis is the index of 64 Pauli operators $\mathcal{P}_i\in \{III, II\sigma_x, ...., \sigma_z\sigma_z\sigma_z\}$ and the vertical axis is the index of 15 measurement settings shown in Table \ref{aaresult}. $A^t_{ij}=\pm 1$ means that the Pauli operator $\mathcal{P}_j$ can be measured in the $i$-th measurement setting.}
\label{aar}
\end{figure*}

\begin{table*}
\caption{The optimal tomographic scheme for the NN configuration. Here, examples of the size with 2 to 4 qubits are presented. }. 
\begin{tabular}{cc}
\toprule[1pt]  
Qubit & ~~~Measurement settings~~~ \\
\hline
2 & $\mathcal{R}^2_x$, $\mathcal{R}^2_y$, $\mathcal{R}^1_x$, $\mathcal{R}^1_y$, $\mathcal{YY}^{(12)}$, $\mathcal{XY}^{(12)}$\\

3 & $\mathcal{R}^2_x$, $\mathcal{R}^2_x\mathcal{R}^3_x$, $\mathcal{R}^2_y$, $\mathcal{R}^2_y\mathcal{R}^3_y$, $\mathcal{R}^1_x\mathcal{R}^3_x$, $\mathcal{R}^1_x\mathcal{R}^3_y$, $\mathcal{R}^1_x\mathcal{R}^2_x$, $\mathcal{R}^1_y\mathcal{R}^3_x$, $\mathcal{R}^1_y\mathcal{R}^3_y$, $\mathcal{R}^1_y\mathcal{R}^2_y$, 
$\mathcal{YY}^{(12)}$, $\mathcal{YY}^{(12)}\mathcal{R}^3_x$, $\mathcal{XY}^{(12)}\mathcal{R}^3_y$, $\mathcal{YY}^{(23)}$,\\
& $\mathcal{R}^1_x\mathcal{XY}^{(23)}$, $\mathcal{R}^1_y\mathcal{YY}^{(23)}$ \\

4 & $\mathcal{R}^2_x\mathcal{R}^4_x$, $\mathcal{R}^2_x\mathcal{R}^4_y$, $\mathcal{R}^2_x\mathcal{R}^3_y$, $\mathcal{R}^2_y\mathcal{R}^4_x$, $\mathcal{R}^2_y\mathcal{R}^4_y$, $\mathcal{R}^2_y\mathcal{R}^3_x$, $\mathcal{R}^1_x\mathcal{R}^4_x$, $\mathcal{R}^1_x\mathcal{R}^4_y$, $\mathcal{R}^1_x\mathcal{R}^3_x$,  $\mathcal{R}^1_x\mathcal{R}^3_y$,  $\mathcal{R}^1_y\mathcal{R}^4_x$, $\mathcal{R}^1_y\mathcal{R}^4_y$, $\mathcal{R}^1_y\mathcal{R}^3_x$, $\mathcal{R}^1_y\mathcal{R}^3_y$,  $\mathcal{YY}^{(12)}\mathcal{R}^4_x$, \\

& $\mathcal{XY}^{(12)}\mathcal{R}^4_x$, $\mathcal{YY}^{(12)}\mathcal{R}^4_y$, $\mathcal{XY}^{(12)}\mathcal{R}^4_y$, $\mathcal{YY}^{(12)}\mathcal{R}^3_x$, $\mathcal{XY}^{(12)}\mathcal{R}^3_x$, $\mathcal{XY}^{(12)}\mathcal{R}^3_x\mathcal{R}^4_x$, $\mathcal{YY}^{(12)}\mathcal{R}^3_x\mathcal{R}^4_y$, $\mathcal{YY}^{(12)}\mathcal{R}^3_y$, $\mathcal{XY}^{(12)}\mathcal{R}^3_y$, $\mathcal{YY}^{(12)}\mathcal{R}^3_y\mathcal{R}^4_x$, \\
& $\mathcal{XY}^{(12)}\mathcal{R}^3_y\mathcal{R}^4_y$, $\mathcal{XY}^{(23)}$, $\mathcal{XY}^{(23)}\mathcal{R}^4_x$, $\mathcal{YY}^{(23)}\mathcal{R}^4_y$, $\mathcal{R}^2_x\mathcal{YY}^{(34)}$, $\mathcal{R}^2_y\mathcal{XY}^{(34)}$, $\mathcal{R}^1_x\mathcal{YY}^{(34)}$, $\mathcal{R}^1_x\mathcal{XY}^{(34)}$, \\
& $\mathcal{R}^1_x\mathcal{R}^2_x\mathcal{YY}^{(34)}$, $\mathcal{R}^1_x\mathcal{R}^2_y\mathcal{XY}^{(34)}$, $\mathcal{R}^1_y\mathcal{YY}^{(34)}$, $\mathcal{R}^1_y\mathcal{XY}^{(34)}$, $\mathcal{R}^1_y\mathcal{R}^2_x\mathcal{XY}^{(34)}$, $\mathcal{R}^1_y\mathcal{R}^2_y\mathcal{YY}^{(34)}$\\
\toprule[1pt]  
\end{tabular}
\label{aaresult2}
\end{table*} 

It is worth stressing that the above result in Eq. (\ref{ssn}) is not the optimal solution to this problem. Next, I solve the binary integer programming optimization via the Gurobi solver to find the optimal result.  For the AA configuration, $|\mathcal{S}^\text{new}_{M}|=3^n+3^{n-2}\cdot 2 \cdot C^2_n$ due to the full connectivity between qubits. As shown in Table \ref{com}, the minimum number of measurement settings is presented as a function of the qubit number. The complexity of 7-qubit QST is reduced to 780 from 2187 with saving over 64\% of measurement settings. In Table \ref{aaresult}, I provide some examples of the  measurement settings to fully reconstruct an $n$-qubit state $(n=2, 3, 4)$.  As shown in Fig. \ref{aar}, I also plot the thin matrix $A^t$ after removing the redundant measurement settings from the matrix $A$.  $A^t$ is a more convenient form for the experimentalist to perform QST and recover the density matrix according to the experimental data, see an example in Section \ref{sec2d}.

\subsection{The NN and 2D Configurations}
\begin{figure*}[htp]
\centering  
\includegraphics[width=0.8\linewidth]{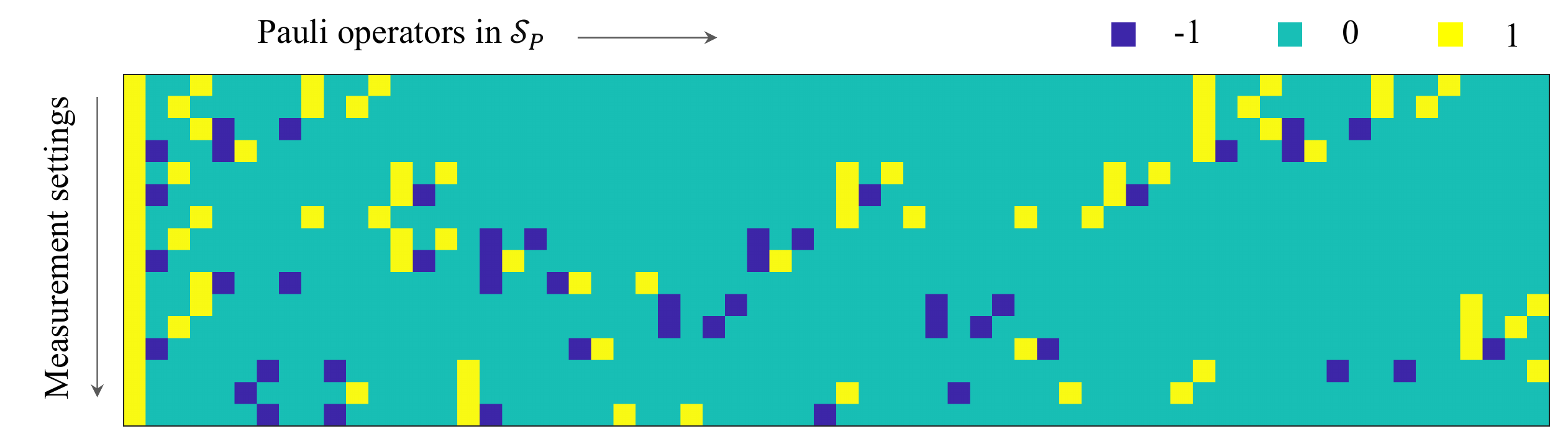}  
\caption{The values of the optimal matrix $A^t$ after removing the redundant measurement settings for a 3-qubit NN configuration. It is a $16\times 64$ matrix, in which the horizontal axis is the index of 64 Pauli operators $\mathcal{P}_i\in \{III, I\sigma_z, ...., \sigma_z\sigma_z\sigma_z\}$ and the vertical axis is the index of 16 measurement settings shown in Table \ref{aaresult2}. $A^t_{ij}=\pm 1$ means that the Pauli operator $\mathcal{P}_j$ can be measured in the $i$-th measurement setting. }
\label{aar2}
\end{figure*}

Considering that the challenge of developing the AA configuration in practice, I further study the optimization of the tomographic scheme for the more common NN and 2D configurations. As shown in Fig. \ref{nn}, the NN connectivity between qubits is a natural arrangement on a linear array of qubits, where two-qubit operations are only available on the nearest-neighbor qubits. Many famous superconducting chips developed on SQC systems adopt this structure. For instance, the 5-qubit and 9-qubit NN superconducting chips arranged in a linear array were developed in Ref. \cite{PhysRevff, barends2014superconducting, kelly2015state}. In Table \ref{com}, I present the optimal number of measurement settings required by QST on an $n$-qubit chain. The corresponding measurement settings are provided in Table \ref{aaresult2}. Figure \ref{aar2} shows the optimal matrix $A^t$ by taking a 3-qubit chain as an example.
The 2D configuration is another chip structure where each qubit can couple with the surrounding qubits, such as the SQC chips from Google and IBM teams \cite{arute2019quantum,wootton2018repetition, wang201816-qubit}, and  a two-by-two planar lattice of SQC qubits \cite{corcoles2015demonstration}. As test examples, I find the minimum number of measurement settings using the integer programming optimization for 4-qubit and 6-qubit 2D structures, and they are also shown in Table \ref{com}. Due to the higher connectivity between qubits, the AA configuration has fewer measurement settings than the NN and 2D configurations to fully reconstruct a quantum state.

\section{Discussion}\label{sec4}
In this section, I further discuss the new scheme and compare it with the traditional one in terms of the feasibility, accuracy, scalability, and measurement.

\textit{Feasibility.}- Let us recall the structure of two-qubit operations $\mathcal{YY}^{(kl)}$ and $\mathcal{XY}^{(kl)}$ described in Eq. (\ref{yyxy}). They include single-qubit rotations $\mathcal{R}^k_{y,z}$ and the couplings evolution $\text{exp}(-i\mathcal{H}_{\text{int}}\tau)$ between $k$-th and $l$-th qubits, and they can be easily implemented on the systems with XX+YY couplings. For instance, $\text{exp}(-i\mathcal{H}_{\text{int}}\tau)$ is a very common $\sqrt{\text{iSWAP}}$ operation on SQC systems,  
\begin{align}
\sqrt{\text{iSWAP}}=\left[\begin{array}{cccc}1 & 0 & 0 & 0 \\0 & 1/\sqrt{2} & -i/\sqrt{2} & 0 \\0 & -i/\sqrt{2} & 1/\sqrt{2} & 0 \\0 & 0 & 0 & 1\end{array}\right]=\text{exp}(-i\mathcal{H}_{\text{int}}\tau).
\end{align}
$\sqrt{\text{iSWAP}}$ operation was proposed and realized a long time ago, which is a natural and easy-to-implemented two-qubit gate using the XY interaction on SQC systems \cite{PhysRevA.67.032301, majer2007coupling, naghiloo2019introduction}. So two-qubit operations $\mathcal{YY}^{(kl)}$ and $\mathcal{XY}^{(kl)}$ can be easily implemented using standard $\sqrt{\text{iSWAP}}$ operations and single-qubits rotations on SQC systems.

\begin{figure}[h]
\centering  
\includegraphics[width=0.78\linewidth]{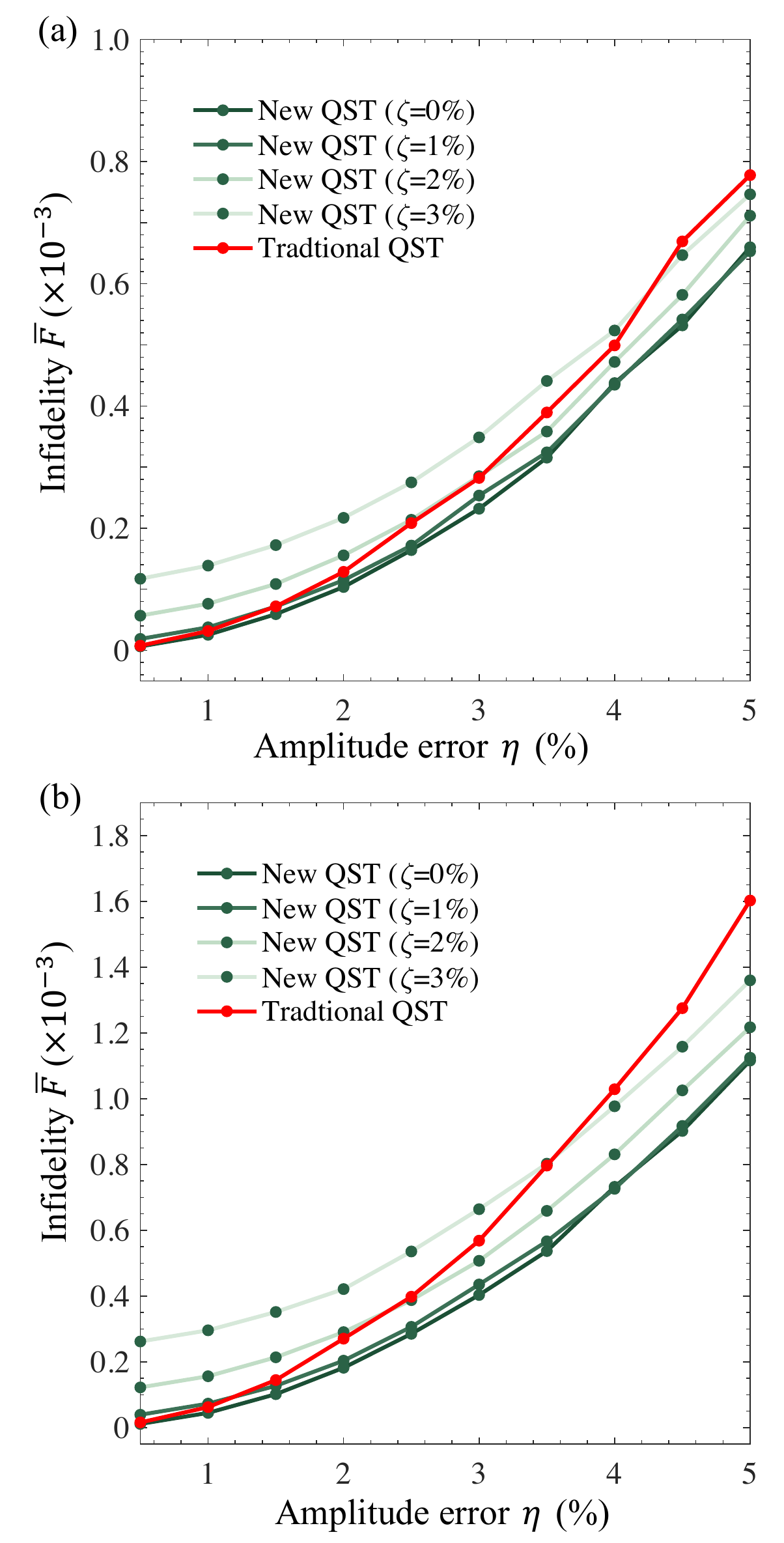}  
\caption{Comparison between traditional QST (red lines) and the proposed scheme (New QST, cyan lines) in the robustness against the control error. (a) 3-qubit states. (b) 5-qubit states. When the residual coupling $\zeta$ is turned off to below 3\%, the new scheme achieves comparable or even better accuracy than the traditional one.} 
\label{sim}
\end{figure}

\textit{Accuracy.}- I numerically simulate the robustness of the new and traditional schemes against the control error in the measurement settings. Here, I mainly consider the amplitude error $\eta$ in single-qubit rotations and the residual coupling $\zeta$ between non-target qubits in two-qubit operations. It is
\begin{align}\nonumber
\mathcal{R}(\theta):& \theta\leftarrow\theta(1\pm u[\eta]),\\
\text{exp}(-i\mathcal{H}_{\text{int}}\tau):& \mathcal{H}_{\text{int}}\leftarrow\mathcal{H}_{\text{int}}+\sum_{\bar{kl}} g_{kl}u[\zeta](\sigma^{\bar{k}}_x\sigma^{\bar{l}}_x+\sigma^{\bar{k}}_y\sigma^{\bar{l}}_y).\nonumber
\end{align}
$u[\eta]$ is a randomized uniform distribution in $[0, \eta]$. In the simulation, $g_{kl}=30 $MHz, $\eta$ is changed from 0\% to 5\%, and $\zeta$ is changed from 0\% to 3\%. I randomly create 2000 $n$-qubit quantum states in the Hilbert space $(n=3, 5)$, and then respectively use the new and traditional QST schemes to reconstruct them. The distance between the ideal states $\rho_0$ and reconstructed states $\rho_e$ is defined by the infidelity $\bar{F}=1-\text{tr}(\rho_0\rho_e)/\sqrt{\text{tr}(\rho_0^2)\text{tr}(\rho_e^2)}$ \cite{fortunato2002design}. Figure \ref{sim} presents the comparison between the new and traditional schemes in the tomography accuracy. The simulation results show that the new scheme achieves comparable or even better accuracy than the traditional one when the residual coupling $\zeta$ between non-target qubits is turned off to below 3\%. To date, the current SQC systems can turn off the residual couplings to below 2\% using the detuning between qubits \cite{barends2014superconducting, kelly2015state, song201710-qubit} or completely turn off the couplings using the tunable coupler \cite{xu2020high, PhysRevApplied.10.054062}.


\textit{Scalability.}- Whether QST has good scalability or not is very important for the larger quantum systems. Indeed, some tomography schemes with the polynomial or even better scalability were proposed, but they are achieved at the cost of focusing on the constrained states. For instance, the tomography with the compressed sensing only works on the states with the low rank \cite{PhysRevLett.105.150401}, the neural network tomography \cite{torlai2018neural, xin2019local-measurement-based} and the Hybrid-Quantum-Classical-based tomography \cite{PhysRevApplied} only aim to reconstruct the ground and dynamical evolved states of many-body Hamiltonians, and the tomography via 2-body reduced density matrices only reconstructs the states with the so-called UD properties \cite{xin2017quantum}. Of course, how to develop the scalable tomography methods for the constrained states is also interesting research, but it is not the topic of this work.


\begin{figure}[h]
\centering  
\includegraphics[width=0.85\linewidth]{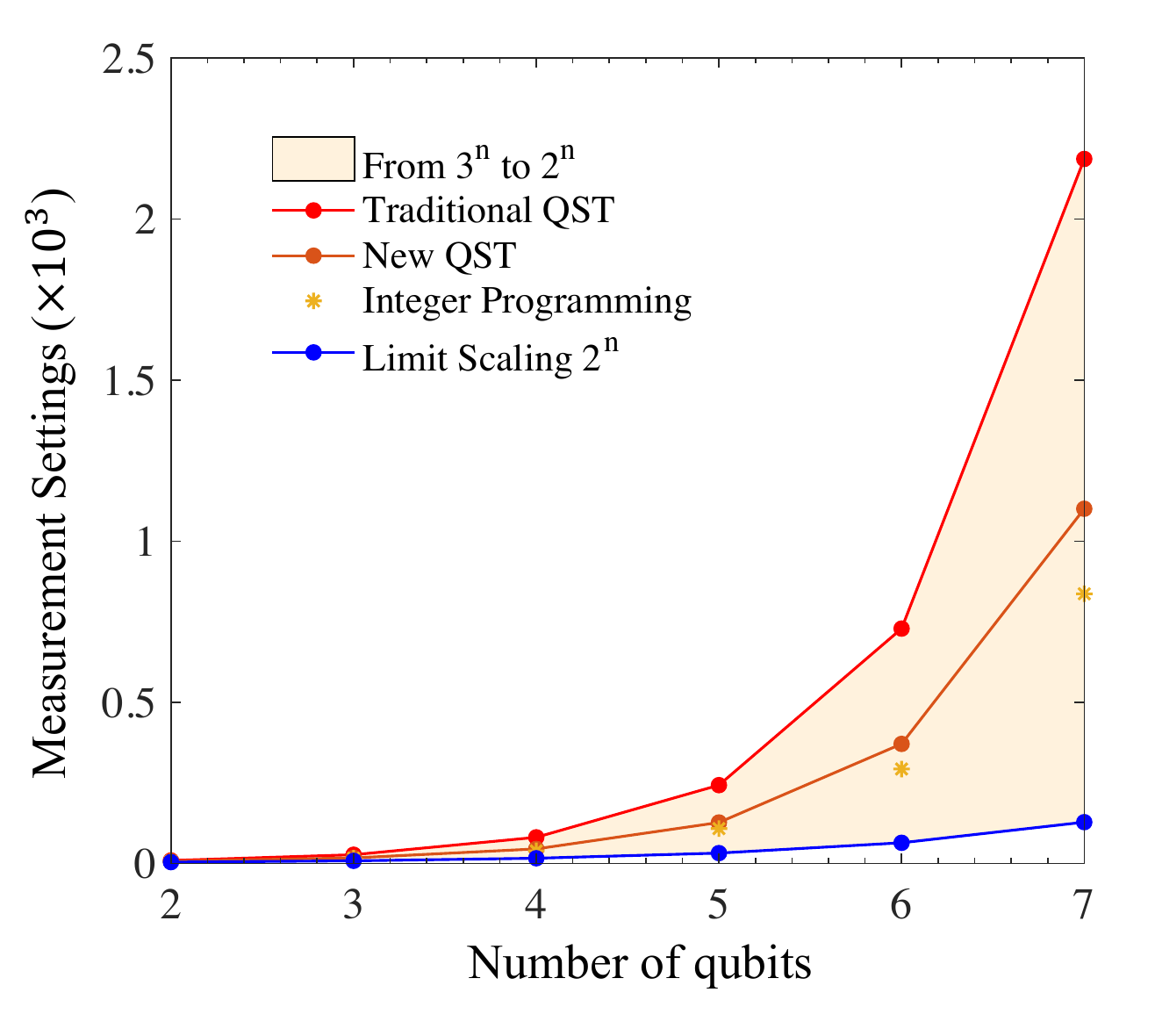}  
\caption{Comparison between the traditional QST and the proposed scheme (New QST and Integer programming) in the number of measurements settings. The red line is the scaling of traditional QST and the blue line is the scaling limit for reconstructing a general state. In principle, the tomography scheme breaking the scaling limit of $2^n$ does not exist for general states in the Hilbert space. } 
\label{sc}
\end{figure}

My work focuses on how to reduce the measurement cost of reconstructing a general state instead of a constrained state in the whole Hilbert space, which is one of the important demands for fully characterizing and developing quantum devices in the Noisy Intermediate-Scale Quantum era. Unfortunately, to the best of my knowledge, it is unlikely to devise a tomography scheme with favorable scaling for a general state due to the intrinsic complexity of the state tomography problem. Take the superconducting circuit as an example, only $2^n$ Pauli operators can be measured by one measurement setting, but an unknown quantum density matrix has $4^n$ Pauli operators to be determined. So at least $4^n/2^n=2^n$ measurement settings are required for fully reconstructing a quantum state. Hence, it is not possible to develop a tomography scheme breaking the exponential scaling. In such a situation, what one can do is to develop tomography schemes that approach the limit of $2^n$ as more as possible. As shown in Fig. \ref{sc}, the traditional QST scheme requires $3^n$ measurement settings, while the new scheme takes a big step from the traditional $3^n$ to the limit $2^n$ with saving up to over 60\% of the measurement settings. So not that the proposed scheme can not achieve a good scaling, but in principle, it does not exist a tomography scheme breaking the exponential scaling when we want to reconstruct a general state instead of a constrained state. The exponential overhead is inevitable to reconstruct generic quantum states of systems. However, the number of measurement settings required by QST can also be significantly reduced for some special states, such as the polynomial scaling for the ground states of local Hamiltonians \cite{PhysRevApplied, xin2019local-measurement-based} and the states with UD property via 2-reduced density matrices \cite{xin2017quantum}.

\textit{Measurement.}- As shown in Fig. \ref{pro}, for an unknown state $\rho$ to be reconstructed, the measurement setting $M_i$ (such as single-qubit rotations or two-qubit operations) is firstly applied on $\rho$  and $\rho_i=M_i\rho M_i^{\dagger}$ is created, and then the expectation value of operator $O_j$ is measured by $\left \langle O_j \right \rangle=\text{Tr}(O_j\cdot \rho_i)$. Here, the measurable operator $O_j$ is independent of the choice of the measurement setting $M_i$. It is an intrinsic property of the quantum devices, such as $O_j \in \{I,\sigma_z\}^{\otimes n}$ on SQC systems  \cite{song201710-qubit} and $O_j \in \{\sigma_x, \sigma_y\}\otimes \{I,\sigma_z\}^{\otimes (n-1)}$ on NMR systems \cite{RevModPhys.76.1037}. Hence, the same as the traditional scheme, the new scheme measures the same operators $O_j$'s after each measurement setting $M_i$.

\section{Conclusion}\label{sec5}

In summary, I proposed a new QST scheme by adding 2-qubit operations on the basis of single-qubit rotation measurement settings and transfer it to the binary integer programming optimization problem to find the optimal tomographic scheme. I also apply the new QST scheme on SQC systems with different configurations between qubits. Compared with the traditional scheme, the new one can save up to over 60\% of measurement settings and it can achieve comparable or even better accuracy when the residual couplings between non-target qubits are turned off to below 3\%. It may be a little hard to use the binary integer programming optimization to find the optimal tomographic scheme for a larger quantum system, but it is still easy to find a suboptimal solution or approximate solution for the intermediate-scale quantum systems. In this work, I assume that one measurement setting includes at most one 2-qubit operation. If we consider more 2-qubit operations or select other elements from the Clifford group, the minimum number of measurement settings will be further reduced. To boost the speed of finding the optimal solution via the integer programming optimization, the symmetry of this problem should be taken into account. For instance, there are some symmetries in the sets $\mathcal{S}_{P}$, $\mathcal{S}_{O}$, and $\mathcal{S}_{M}$. If these symmetries are considered in solving the integer programming optimization, the computation amount of reaching the optimal solution is likely reduced \cite{margot2010symmetry}, which is an interesting question for the tomography of the larger systems in the future research. The results obtained in this work can be applied to quantum platforms including but not limited to SQC systems, and the ideas and methods developed in this work will be also  helpful in designing the feasible tomographic experiments in practice.

\begin{acknowledgments}
This work was supported by the National Key Research and Development Program of China (Grant No. 2019YFA0308100), the National Natural Science Foundation of China (Grants No. 11975117, No. 11875159, No. 11905099,  and No. U1801661), Science, Technology and Innovation Commission of Shenzhen
Municipality (Grants No. JCYJ20180302174036418), Guangdong Basic and Applied Basic Research Foundation (Grant No. 2019A1515011383), and Guangdong Provincial Key Laboratory (Grant No. 2019B121203002).


\end{acknowledgments}


%

\end{document}